\documentclass[aip,jcp,reprint,amssymb,longbibliography]{revtex4-1}

\usepackage{graphicx}
\usepackage{hyperref}
\usepackage{amsfonts}
\usepackage{color}
\usepackage{amsmath}
\usepackage{listings}
\usepackage[T1]{fontenc}
\usepackage[utf8]{inputenc}
\usepackage[above,below]{placeins}
\usepackage{newfloat}

\usepackage{algpseudocode} 	
\usepackage{algcompatible}	

\usepackage[normalem]{ulem}	

\makeatletter
\let\l@English\relax
\makeatother

\DeclareFloatingEnvironment[
    fileext=loa,
    listname=List of Algorithms,
    name=ALGORITHM,
    placement=tbhp,
]{algorithm}

\begin{document}

\def\dr{{\,\rm d}{\bf r}}
\def\E{{\bf E}}
\def\P{{\bf P}}
\def\mm{{\bf m}}
\def\D{{\bf D}}
\def\r{{\bf r}}
\def\curl{{{\rm curl}\; }}
\def\grad{{{\rm grad}\; }}
\def\div{{{\rm div}\; }}
\def\p{{\bf p}}
\def\rhat{\hat{\bf r}}
\def\dq{ \, {\rm d}q\, }
\def\ss{{\rho_f}} 
\def\q{{\bf q}} 
\def\P{{\bf P}} 
\def\ash{{{\rm      sinh^{-1}} }} 
\def\L{{\mathcal{L}}} 
\def\tphi{{\tilde{\phi}}}
\def\tF{{\tilde{F}}} 
\def\trho{{\tilde{\rho}}} 
\def\tD{{\tilde{\D}}}
\def\tmu{{\tilde{\mu}}} 
\def\tE{{\tilde{\E}}} 
\def\tf{{\tilde{f}}}
\def\C{{\mathcal{C}}}

\title{Convexity and stiffness in energy functions for electrostatic simulations}
\author{Justine S. Pujos, A.C. Maggs} \affiliation{Laboratoire PCT,
  Gulliver CNRS-ESPCI UMR 7083, 10 rue Vauquelin,75231 Paris Cedex 05,
  France} \date{\today}
\begin{abstract}
  We study the properties of convex functionals which have been
  proposed for the simulation of charged molecular systems within the
  Poisson-Boltzmann approximation. We consider the extent to which the
  functionals reproduce the true fluctuations of electrolytes and thus
  the one-loop correction to mean field theory -- including the
  Deby-H\"uckel correction to the free energy of ionic solutions. We
  also compare the functionals for use in numerical optimization of a
  mean field model of a charged polymer and show that different
  functionals have very different stiffnesses leading to substantial
  differences in accuracy and speed.
\end{abstract}
\maketitle

\section*{Introduction}

The sign of electrostatic free energy functionals has long interested
and puzzled the community of molecular simulators: A recent paper states {\it
  The fact that the functional cannot be identified with the
  electrostatic energy away from the minimum a priori precludes its
  use in a dynamical ``on the fly''
  optimization\dots}\cite{allen}. Similar statements as to the nature of
the free energy in Poisson-Boltzmann functionals can be found in other
papers\cite{PB02,reiner}.

Implicitly most formulations of electrostatic free energies
work\footnote{an exception is~\cite{vanderbilt} } with the electric
field and the potential and are associated with the natural potential
energy
\begin{equation}
  U_E = -\int \frac{\epsilon \E^2}{2} \dr \label{eq:concave}
\end{equation}
This energy seems to be concave and unbounded below, whereas when
thermodynamic potentials are written in terms of the electric
displacement field $\D$ we find
\begin{equation}
  U_D = + \int \frac{\D^2}{2 \epsilon}\dr
\end{equation}
which is convex.  As emphasised in a classic text\cite{landau} the
two formulations are identical in content, and simply linked by a
Legendre transform. The importance of the correct ensemble in the
understanding of electrodynamics and stability of media was
particularly emphasised in reviews of Kirzhnits and collaborators
\cite{kirzh,sign}. Inspired in particular by this view we proposed a
free energy functional for the Poisson-Boltzmann equation which is a
true minimizer\cite{Tony-LT, pujos2012legendre}, rather than a
stationary principle.  Our hope is that when implemented in practical
codes\cite{holm} these locally formulated convex forms give more
stable and simpler algorithms. Using a convex free energy it is
possible to implement an implicit solvent code which fully includes
the Poisson-Boltzmann free energy in which ``on the fly'' optimization
can then be performed using the Car-Parrinello method\cite{car}. In
this case the electrodynamic field can evolve in parallel to degrees
of freedom in the molecule \cite{rottler}.

Our work takes the displacement field $\D$ as the fundamental
thermodynamic field, rather than $\phi$ or $\E$. Other recent work
\cite{PB01-Monica, monica2, monica3} takes a different approach to the
problem. By a series of transformations to the original variational
formulations of the Poisson-Boltzmann functional (expressed as a
function of the electrostatics potential, $\phi$) the authors have
found new families of functionals which are scalar, convex and local.

While all these functionals are strictly equivalent at their
stationary point when simulating at finite temperatures differences
can occur: The Car-Parrinello method thermostats the supplementary
variables to zero temperature; in this way we only see the true
minimum of the energy functional. However in many applications we
might wish to run simulations with {\it all variables thermostated to
  the same, ambient temperature}. For electrostatic functionals in
dielectric media this samples fluctuations which are equivalent to the
so-called thermal Casimir interaction\cite{auxiliary,dynamics,sam},
which is just the zero frequency contribution to the free energy in
Lifshitz theory. There thus arises the question -- what happens to the
free energies when a convex Poisson-Boltzmann functional is used in a
finite temperature simulation? One might hope that the functionals
also reproduce the correct one-loop corrections to the free energy,
which in the case of an electrolyte has an anomalous scaling in
$\rho^{3/2}$ where $\rho$ is the salt concentration.

To answer this question we will study the spectra of several free
energy functionals expanded to quadratic order. The spectra are also
crucial in understanding the convergence properties and dynamics of
simulation algorithms. Accurate (non-implicit) integration in
molecular dynamics requires a time step which is short compared with
all dynamic processes under consideration. Algorithms which generate
time scales which are very different for different modes are said to
be {\it stiff}\/ and lead to reduced efficiency. Stiffness can also
have other, unwanted consequences in optimization algorithms -- such
as poor convergence properties. We explore this question in the last
part of the paper where we study a model of free energy minimization
in the context of a confined polyelectrolyte.

We start by a reminder of Poisson-Boltzmann theory and the corrections
to it coming from Gaussian fluctuations at the one-loop level. We then
show that the same one-loop free energy is found in a complementary
dual (and convex) formulation of the free energy which can be used
within a full molecular dynamics or Monte Carlo simulation.  We then
move on to considerations of coarse grained models and test various
discretizations of electrostatics energies for their accuracy and ease
of use within minimizing principles.

\section*{Poisson-Boltzmann theory and its one-loop correction}
In our presentation we will consider the case of a symmetric
electrolyte, however the identities that we derive are independent of
the exact microscopic model used. The use of a definite physical
example leads to considerable simplification of notation.

The mean-field Poisson-Boltzmann functional for a symmetric
electrolyte is
\begin{equation}
  F_{\phi}=\int \left( - \frac{\epsilon (\nabla \phi)^2}{2} - 2 k_B T
    \lambda
    \cosh{\beta e \phi} + \rho_e \phi \right ) \dr \label{eq:free}
\end{equation}
where $k_BT=\beta^{-1}$ is the thermal energy, $e$ the ion charge
$\phi$ the electrostatic potential and $\lambda$ a fugacity.  At the
stationary point of this ``free energy'' we find
\begin{equation}
  \div \epsilon \grad \phi - 2 \lambda  e \sinh(\beta e \phi) +
  \rho_e=0 \label{eq:pb}
\end{equation}
which is indeed the classic equation for the potential in the
Poisson-Boltzmann formalism. Linearizing this equation allows us to
define the inverse Debye length from
$\kappa^2=2\lambda \beta e^2/\epsilon$. At low concentrations
$\lambda$ can be identified with the salt concentration.  Since this
thermodynamic potential is naturally expressed in terms of the
thermodynamics fields $\phi$ or $\E$ it is concave as noted for
eq.~(\ref{eq:concave}).  Any attempt to use such a form in
Car-Parrinello simulation will lead to numerical instabilities.

Much recent work has used field theory techniques to extend this
functional to include fluctuations. In particular one can sum the
fluctuations to the {\it one-loop}\/ level \cite{zeks,henri}, which
gives as a correction to this free energy the functional determinant
\begin{equation}
  F_{loop} = \frac{k_BT}{2} \log \big \lvert -\div \epsilon(r) \grad + 2
  \lambda \beta e^2 \cosh{(\beta e \phi)}  \big \rvert \label{eq:loop}
\end{equation}
This expression is often interpreted with a substraction scheme -- one
compares with the free energy of an empty box, where $\lambda=0$. Even
with this subtraction the expression is formally divergent. However on
introducing a microscopic cut-off\cite{wang} the loop free-energy
gives both the Born self-energy of solvation
\begin{equation}
  E_{Born} = \frac{e^2}{8 \pi \epsilon a}
\end{equation}
as well as the anomalous Debye-H\"uckel free energy
\begin{equation}
  \frac{F_{DH}}{V} =- \frac{k_BT  \kappa^3}{12 \pi}
\end{equation}
as is shown in Appendix II.

\section*{Properties of the Legendre transform}
As noted above many of the standard concave expression for
electrostatic free energies can be rendered convex by Legendre
transform \cite{Tony-LT}. Several conventions exist in the
literature\cite{zia}. In particular we define the transform of a
convex function $c(x)$ as
\begin{equation}
  g(s) = \sup_x ( s x -c(x) )
\end{equation}
We will also use the notation $g(s) = \mathcal{L}[c](s)$.  The
transformation is an {\it involution}, that is the double transform of
a convex function is the identity.

For this paper we will be interested in the second order
expansion of free energies about an equilibrium point. For this we
will use an important identity linking the second derivatives of the
functions $g$ and $c$.
\begin{equation}
  \frac{d^2 c}{d x^2} \frac{d^2 g}{d s^2}=1 \label{eq:curve}
\end{equation} 
valid at the corresponding pair of variables $(x,s)$. We also note
that the Legendre transform of a scaled function:
$\alpha c( \gamma x)$ is given by $\alpha g( s/(\alpha \gamma))$.

\section*{Dual formulations of the Poisson-Boltzmann equation}
In previous work \cite{Tony-LT, pujos2012legendre} we have detailed
how to pass from the Poisson-Boltzmann free energy expressed in terms
of the electrostatic potential, $\phi$ to a dual formulation expressed in
terms of the electric displacement $\D$. In particular we found
\begin{equation}
F_{dual} = F_{field} + F_{ions}
\end{equation}
where
\begin{equation}
F_{field} = {\mathcal L} [\epsilon \E^2/2] = \frac{\D^2}{2 \epsilon}
\end{equation}
with $\E= - \grad \phi$.
and
\begin{equation}
F_{ions} = {\mathcal L} [2\lambda k_BT \cosh(e \beta \phi)] (\div \D
-\rho_e) 
\end{equation}
The transform of the hyperbolic cosine is easily found
\begin{equation}
 g(s)= {\mathcal L}[2\cosh](s) =s \sinh^{-1} (s/2) -\sqrt{4+s^2}  
\end{equation}
thus
\begin{equation}
F_{dual} = \int \left ( \frac{\D^2}{2 \epsilon} + k_BT \lambda  g
  \left ( \left (\div \D
  -\rho_e \right )/e \lambda  \right ) \right )
\dr \label{eq:dual}
\end{equation}
This form as a sum of two Legendre transforms will turn out to be very
important for a number of determinant identities that we derive below.
We will need also the expansion of $g(s) = -2 +s^2/4 -s^4/192 +\dots$.

This transformation gives a convex free energy which can be used in a
Monte Carlo or molecular dynamics simulation. It has been constructed
so that the minimum of eq.~(\ref{eq:dual}) is identical to the maximum
of eq.~(\ref{eq:free}). The potential and dual formulations  are linked via
the equation:
\begin{equation}
\div \D - \rho_e = 2 \lambda e \sinh{(\beta e \phi)} \label{eq:link}
\end{equation}
which can be understood as showing that the ionic charge concentration
in the Poisson-Boltzmann equation is $2 \lambda e \sinh{\beta e \phi} $.

\subsection*{Parameterization and Scalarizing}
It is clear that the use of a vector functional rather than the usual
scalar form has increased the number of variables that need to be
simulated or optimized\cite{monica2,monica3}. We show here that given
a functional eq.~(\ref{eq:dual}) of a vector field we can find a
related functional of a scalar field.

When we consider the stationary point of the energy
eq.~(\ref{eq:dual}) we find that
\begin{equation}
  \frac{\D}{\epsilon} - \frac {k_B T } {e } \grad g' =0
\end{equation}
We then identify the function $k_BT g'/e = - \phi$ at the
stationary point. We see that we can make such a substitution even
away from the minimum to find a more restrictive functional which is
to be optimized over a smaller subspace:
\begin{equation}
  F_{dual,\phi} = \int \Big (\epsilon \frac{(\nabla \phi)^2 }{2 } +
  k_BT \lambda g( (\div \epsilon\grad \phi
  +\rho_e)/e \lambda  ) \Big ) \dr \label{eq:ourscalar}
\end{equation}
Since the space is more restrictive the minimum of this functional is
clearly greater than or equal to the functional expressed in terms of
$\D$, however the absolute minimum is compatible with the
parametrization, implying that this functional has the correct minimum
free energy. Indeed we can go further and take linear combinations
\begin{equation}
  F_{comb} = (m+1 )F_{dual,\phi} + m F_{\phi}
\end{equation}
for $m \ge 0$ which all have the same, correct minimum.

\subsection*{Determinant identities}

In this section we present some matrix identities which link the
functional determinants of operators with their dual equivalent. These
identities are particularly interesting because they link operators
(at least when discretised) of dimensions $N\times N$ which are
expressed in terms of the potential, to operators of dimensions
$3N\times 3N$ for the electric displacement.

The identities are easiest to derive from a discretized form of the
free energy. We will work with the discrete divergence operator
$\partial_{rl}$ which acts on the link variable $D_l$. When
discretizing to a simple cubic lattice there are $3N$ variables $\D_l$
and $N$ values of the divergence which we associate with the lattice
points. We discretize the non-linear contribution to the free energy
eq.~(\ref{eq:dual}) as follows:
\begin{equation}
  F_{ions} =  \sum_r g  \left  ( {\textstyle \sum_l } \partial_{rl} D_l - \rho_e \right  )
\end{equation}
Take a second derivative with respect to the link variables $D_p$ and
$D_q$ to find the matrix
\begin{equation}
  \Delta_{pq}= \sum_r      \partial_{rp} \partial_{rq}
  g'' \left ( \textstyle \sum_l \partial_{rl} D_l  - \rho_e \right  ) \label{eq:gpp}
\end{equation}
where $g''$ is the second derivative of $g$, evaluated at $r$.  Create a matrix with
$g''$ on the diagonal then eq.~(\ref{eq:gpp}) is a conventional matrix
product:
\begin{equation}
  \Delta_{pq}=  \sum_{rr'}\partial^{T}_{pr}  (g_r'' \delta_{rr'}) \partial_{r'q}
\end{equation}
We identify $(-\partial^T_{pr})$ with the gradient operator since
$\div$ and $(-\grad)$ are mutually adjoint.

We similarly discretize the gradient contribution to the free energy
eq.~(\ref{eq:free})
\begin{equation}
  \frac{1}{2}  \sum_{rr',l}  \epsilon_l  ( \partial^T_{l r}  \phi_r )  (\partial^T_{l r'}
  \phi_r' )
\end{equation}
and take the second derivative with respect to $\phi_r$ to find the
matrix
\begin{equation}
  \tilde \Delta_{rr'} =  \sum_l \epsilon_l  \partial^T_{l r} \partial^T_{l r'}
\end{equation}
When we turn $\epsilon_l$ into a diagonal matrix we find the matrix
product:
\begin{equation}
  \tilde \Delta_{rr'} =  \sum_{ll'} \partial_{r'l} (\epsilon_l \delta_{l l'}) \partial^T_{l' r}
\end{equation}
which is the discrete version of the operator
$(-\div \epsilon \grad)$.

We now consider the relation between the one-loop free energy,
eq.~(\ref{eq:loop}), and its naive equivalent within the dual
formulation found by taking the second functional derivative of
eq.~(\ref{eq:dual}) which we write as the discretized form:
\begin{equation}
  F_{dual,loop} =  \frac{k_BT}{2} \Big |   \frac{\delta_{ll'}
  }{\epsilon_l }   + \partial^T g'' \partial
  \Big | \label{eq:dualloop}
\end{equation}
We will show that modulo some trivial local contributions to the free
energy the expression eq.~(\ref{eq:dualloop}) contains the same
physics as eq.~(\ref{eq:loop}) which we write as
\begin{equation}
  F_{loop} =  \frac{k_BT}{2} \Big |
  \partial \epsilon \partial^{T} + c'' 
  \Big | 
\end{equation}
where $c''$ is the second derivative of the entropic contribution to
the free energy. In the case of the symmetric electrolyte
$c= 2 k_BT \lambda \cosh(e \beta \phi)$

To link these two determinants we will make use of the singular value
decomposition theorem which states that a {\it rectangular matrix}\/,
$A$ of dimensions $m\times n$ can be expressed as a product
\begin{equation}
  A = U \Sigma V^*
\end{equation}
where $U$ is an unitary matrix of dimensions $m\times m$ such that
$UU^*=U^*U=1$; $V$ is also unitary of dimensions $n\times n$ and
$\Sigma$ is diagonal rectangular with non-negative elements on the
diagonal.

We consider the determinant for a rectangular matrix $A$:
\begin{equation}
  \big | A  A^T + 1 \big |
\end{equation}
With the help of the singular value decomposition we can write
\begin{align}
  \big | 1 + A A^T \big | = &\big |1+U \Sigma \Sigma^T U^* \big | =
                              \big |1 + \Sigma \Sigma^T \big | \\ = &
                                                                      \big |1+
                                                                      \Sigma^T \Sigma \big | = \big | 1 + A^T A \big |
\end{align}
The identity is between two matrices of different dimensions:
$m\times m$ and $n\times n$.  We apply this theorem to
\begin{equation}
  A = \frac{1}{\sqrt{c''}} {\rm \partial \, } \sqrt{\epsilon}
\end{equation}
and find
\begin{equation}
  \Big| 
  \frac{1}{\sqrt{c''}} {\partial\, }{\epsilon}\partial^T  \frac{1}{\sqrt{c''}} +1 
  \Big |
  =
  \Big |
  \sqrt{\epsilon}  \partial^T \frac{1}{c''} \partial \sqrt{\epsilon} +1
  \Big |
\end{equation}
We can pull the diagonal matrices $c''$ and $\epsilon$ out of the main
determinants to find
\begin{equation}
  \Big |
  \partial \epsilon \partial^T + c''
  \Big  |
  =
  \big |c'' \big | \big |\epsilon \big |
  \Big |
  \partial^T \frac{1}{c''} \partial +\frac{1}{\epsilon}
  \Big |
\end{equation}
We now make use of the relation between the curvatures of a function
and its Legendre transform eq.~(\ref{eq:curve}): $g''=1/c''$, so that
\begin{equation}
  \Big |
  \partial \epsilon \partial^T + c''
  \Big  |
  =
  \big |c'' \big | \big |\epsilon \big |
  \Big |
  \partial^T g'' \partial +\frac{1}{\epsilon}
  \Big |
\end{equation}
which is our required identity. The result is rather remarkable. The
dual energy eq.~(\ref{eq:dual}) was derived purely by considering
stationnary properties of the free energy. {\it However}, if we use
this in a simulation at finite temperature {\it we generate
  automatically the correct one loop free energy}, modulo a trivial
shift in the zero. Of course we do not have any guarantee that any higher
order contributions are correct -- in particular in the strong
coupling limit\cite{sahin,sahin2,rudireview} the addition of terms valid in
a weak coupling expansion may not help in correctly describing the
physics.  We can also write the dual contribution to the free energy
as
\begin{equation}
  F_{dual,loop}= \frac{k_BT}{2} \log \Big |
  - \grad g'' \div + \frac{1}{\epsilon}
  \Big |
\end{equation}
For the specific case of a symmetric electrolyte
$d^2g/ds^2 = 1/\sqrt{4+s^2}$.

As noted in a recent paper the one-loop corrected Poisson-Boltzmann
equation contains many interesting physical effects including image
charges in the case of dielectric continuities\cite{wang,bPB04}. Thus
having a form of the Poisson-Boltzmann that can be simulated and that
includes such terms could be particularly interesting in applications
near surfaces.

\section*{General scalar functionals} 
Entirely different approaches to convexification have also been
proposed in a recent series of papers\cite{PB01-Monica, monica2,
  monica3}. In particular it was shown that it was possible to develop
a local, convex functional which is expressed in terms of the scalar
potential, rather than the vector field $\D$. The authors give several
expressions but in particular find a local minimizing principle with
the functional
\begin{align}
  F_{sc}=& \int 
           \Big [  \frac{\epsilon (\nabla \phi)^2 }{2} 
           - {2 k_BT\lambda} \cosh{(\beta e \phi)}\nonumber
  \\+& 2 \lambda e \phi \sinh(\beta e \phi) \\ \nonumber  
  +&\frac{k_BT}{e}\sinh^{-1} \left ( (\div \epsilon \grad \phi +
     \rho_e ) /2e\lambda \right ) \times \nonumber  \\ & 
                                                         (  \div \epsilon \grad \phi +  \rho_e -2 e \lambda \sinh(\beta e \phi) ) \Big ] \dr \label{eq:monica}
\end{align}
When expanded to second order in $\phi$ this gives:
\begin{align}
  F_{sc} = {\rm const}+ \int \Big [
  &\frac{3 \epsilon (\nabla \phi)^2 }{2} 
    +\lambda \beta   e^2 \phi^2 \nonumber \\+&
                                     \frac {(\div \epsilon \grad \phi + \rho_e)^2}{2 e^2 \beta \lambda} 
                                     - \rho_e \phi \Big ]
                                     \dr \label{eq:comb}
\end{align}
At least to second order this looks like a linear combination of
eq.~(\ref{eq:ourscalar}) and eq.~(\ref{eq:free}).  Far from any
sources in a background and with uniform dielectric properties we can
re-write eq.~(\ref{eq:comb}) in Fourier space:
\begin{equation}
  F=\sum_q  \Big ( \frac {3\epsilon q^2}{2} + \lambda \beta e^2  +\frac{\epsilon^2q^4 }{2 \lambda
    e^2 \beta}  \Big ) \lvert \phi_q \rvert^2 \label{eq:linmoneca}
\end{equation}

The question that arises at this moment is whether such an expression
also gives rise to the correct one-loop potential. From the dispersion
relation we deduce that the fluctuation potential can be written in
the form
\begin{equation}
  F_{fluc} =\frac{k_BT}{2} \sum_q \log \left (1 + 3 (q /\kappa)^2 +(q /\kappa)^4
  \right )
\end{equation}
which corresponds to eq.~(\ref{eq:oneint}) in Debye-H\"uckel theory.
As can be expected this gives rise to a divergent integral in three
dimensions. However it is not clear how to separate this expression
into a Born energy with a Debye-H\"uckel fluctuation potential of the
form given in Appendix II, eq.~(\ref{eq:debyeIntegral}). We must
regularize by subtracting off $\log(q^4)$ but this still leaves a
Born-like energy which has a radius which is itself a function of the
Debye length. Our conclusion is that the scalar functionals are indeed
correct for the ground state, but do not correspond to the correct
excitation spectrum of the true physical system. These functionals can
be used in a Car-Parrinello scheme where they are thermostated to zero
temperature to avoid fluctuations.

\section*{Spectrum and stiffness}

In a uniform dielectric background far from charges the determinant
eq.~(\ref{eq:loop}) can be diagonalised using a Fourier decomposition. The eigenvalues of
the operator in eq.~(\ref{eq:loop}) are then
\begin{equation}
  \omega_q =  \epsilon ( q^2 + \kappa^2) \label{eq:disp1}
\end{equation}

We now consider the spectrum in the dual picture by expanding the
free energy eq.~(\ref{eq:dual}) to quadratic order, to find the vector
equivalent of the Debye-H\"uckel theory.
\begin{equation}
  F_{vec} = {\rm const}+ \int \left [\frac{\D^2}{2 \epsilon} + k_BT  \frac{(\div \D -\rho_e)^2}{4
      \lambda e^2} \right ] \dr
\end{equation}
The spectrum for quadratic fluctuations decomposes into a longitudinal
and transverse part.
\begin{align}
  \omega_l=& \left (\frac{1}{\epsilon} + \frac{k_BT q^2}{2 \lambda
             e^2} \right ) = \omega_0 (q^2+ \kappa^2) \label{eq:disp2}\\
  \omega_t=&\frac{1}{\epsilon} \nonumber 
\end{align}
We note that the functional forms of eq.~(\ref{eq:disp1}) and
eq.~(\ref{eq:disp2}) are the same which explains, at least partly, the
identity between the free energies at the one-loop level.

Consider now discretizing to a grid of step $h$, finer than the Debye
length. There is a gap in the spectrum eq.~(\ref{eq:disp2}) and the
ratio of the stiffest to the softest modes scales as
\begin{equation}
  S \sim (1/\kappa h)^2 \label{eq:stiff}
\end{equation}
It is clear that this ratio can become large when using a very fine
grid, in which case the system of equations describing the electrolyte
can become stiff.  When integrating a system with a second order
algorithm in time (such as molecular dynamics) the number of time
steps required to sample all modes in the system scales as
$\sqrt{S}$. In a Monte Carlo simulation we can expect that the
equilibration of the electrolyte takes $O(S)$ sweeps. Stiffness slows
the convergence of both molecular dynamics and Monte Carlo algorithms.

If we make a similar calculation for energy functionals such as
eq.~(\ref{eq:linmoneca}) we find that the equations are stiffer: The
spectrum can be written in the form
\begin{equation}
  \omega  = \omega_0 \left (1+ 3(q/\kappa )^2 + (q/\kappa)^4 \right ) \label{eq:spect3}
\end{equation}
 which implies that on a fine grid
\begin{math}
  S \sim (1 /\kappa h)^4
\end{math}
For eq.~(\ref{eq:ourscalar}) the stiffness is given by
\begin{math}
  S \sim  L^2/\kappa^2 h^4
\end{math}
where $L$ is the system size, The presence of higher derivatives in a
functional can imply a slower algorithm, even though fewer degrees of
freedom must be integrated.

We now move on to the second major topic of this paper which is the
performance of optimization algorithms that look for minima in a free
energy.

\section*{Searching for saddle points}


We start by noting that the minimization of a convex function is
generally easy. Many algorithms can be proved to converge quickly
to the correct point. As an example for general quadratic minima the
conjugate gradient method for $N$ variables converges in $N$
iterations. If the object being minimized is a local free energy each
evaluation of the energy is itself of complexity $O(N)$. Thus we
expect that the solution time for a model discretized to $N$ lattice
points should scale as $O(N^2)$. We note that this may not be the
optimal method of solution -- sophisticated solvers of the
Poisson-Boltzmann equation such as Aquasol are rather based on
multigrid methods\cite{koehl,PB03}.

When searching for saddle points no such simple results can be
found. Consider for instance a simple alternating scheme applied to
two variables $x$ and $y$ with the energy
\[ f = - x^2/2 + y^2/2 + \alpha x y\]
For all \(\alpha\) the saddle point remains at the origin.

A simple algorithm is to alternate the two variables and optimized at
each step giving the maps:
\begin{align} x \rightarrow \alpha y \quad y\rightarrow y \\ y
  \rightarrow - \alpha x \quad x \rightarrow x\end{align} A cycle of
updates gives the product
\[
\begin{pmatrix} x'\\y' \end{pmatrix} =
\begin{pmatrix} 0&\alpha\\0&-\alpha^2 \end{pmatrix} \begin{pmatrix}
  x\\y \end{pmatrix} \]
This matrix has eigenvalues \(0, \alpha^2\).
We converge to the origin only when \(\alpha^2<1\).
We learn that mixed optimization over coupled convex-concave spaces
can fail when the coupling between variables becomes too strong in a
way which is impossible for simple convex optimization.

Such mixed saddle point optimization problems occur very often in
electrostatic problems which are formulated in terms of the
electrostatic potential coupled to some other, physical degree of
freedom. In the field of molecular simulation we could consider the
configuration of a macromolecule described by atomistic potentials. As
a more definite example we will consider a model of a charged
polyelectrolyte, in which a long polymer is
described using an auxiliary field $\Psi$, where $\Psi^2$ describes
the local monomer concentration\cite{PolyEl08,PolyEl07-PGG}. We now
consider a number of optimization techniques applied to such a
system. The techniques that we try are
\begin{itemize}
\item Direct nested optimization between the polymer and
  electrostatic potential
\item Building a positive functional from squared derivatives
\item Legendre transform of the electrostatic free energy
\item Use of generalized scalar functionals for the electrostatics
\end{itemize}
We evaluate the methods as a function of the discretization looking
for methods that are rapid and converge easily without extra
derivative information being given by the user.
\subsection*{A model for mixed optimization in electrostatics}

We work with a model developed to study spontaneous assembly of
single-stranded RNA viruses\cite{Vir01,Vir02}.  The genetic material of the virus is
modeled by a single polyelectrolyte chain, which must insert itself into
a charged capsid. The system is supplemented by a symmetric
electrolytic solution. The free energy density describing the
interaction between the polymer, capsid and the electrolyte is
\begin{align}
    F( \phi, \Psi ) = & \int \Big [ - \frac{\epsilon }{2}
    (\nabla\phi)^2
    +   \phi \left( \sigma - p e \Psi^2 \right)\nonumber  \\
    & \; - 2 \lambda \, k_B T  \cosh \left( \beta e \phi
      \right)\nonumber  \\
    & \; + k_B T \left\lbrace \frac{a^2}{6} (\nabla\Psi)^2 +
      \frac{v}{2}\Psi^4 \right\rbrace \Big ] \dr \:.
  \label{eq:baseF}
\end{align}
The third line of this expression corresponds to the non-electrostatic
interactions of the
polyelectrolyte. $\sigma$ is the fixed charge of the capsid, $p$ the
charge on a monomer of the polyelectrolyte, $a$ the monomer size and
$v$ the excluded volume of the polyelectrolyte chain.  This free
energy is concave with respect to $\phi$ and convex with respect to
$\Psi$. These two fields interact through the coupling
$p e \Psi^2 \phi$.  In our studies we only consider systems with
spherical symmetry, discretizing to $N$ points, see Appendix I.

\subsubsection*{Nested Optimization}

The stationary point of the functional $F$ in
equation~(\ref{eq:baseF}) corresponds to the maximum over $\phi$ and
the minimum over $\Psi$.  The most direct way to find such a point is
to search independently for the two extrema: one optimizes iteratively
the configuration while solving the electrostatic problem at each
calculation of the iterative method. The outside loop of optimization
calls upon the interior one until the saddle point is reached (see the
algorithm (\ref{pgm:DL}) below).  We call this method nested
optimization. It is the kind of algorithm which is easy to implement if
one has a pre-existing Poisson-Boltzmann solver which one wishes to
use to study the relaxation of a bio-molecular system without
reformulating the energy functionals.

\begin{algorithm}
\caption{Nested Optimization Program}
\label{pgm:DL}
  \begin{algorithmic}
    \State \textbf{input} \State $\phi \gets \phi_0$
    \State $\Psi \gets \Psi_0$ \\
    \Loop  $\quad$ over $\Psi$ \\
    \Loop $\quad$ over $\phi$ for $\Psi_{fixed} \gets \Psi$ \State
    minimize $ -F(\phi,\Psi_{fixed}) $ over $\phi$ \State
    \textbf{return} $ \phi_{opt} $, $ F( \phi_{opt},\Psi_{fixed}) $
    \State $\phi \gets \phi_{opt}$
    \EndLoop \\
    \State $\phi_{fixed} \gets \phi$ \State minimize
    $ F(\phi_{fixed},\Psi) $ over $\Psi$ \State \textbf{return}
    $ \Psi_{opt}$, $ f(\phi_{opt},\Psi_{opt}) $ \State
    $\Psi \gets \Psi_{opt}$
    \EndLoop \\
    \State \textbf{output} $ \Psi_{OPT} $, $ \phi_{OPT} $,
    $ F( \phi_{OPT} ,\Psi_{OPT}) $
  \end{algorithmic}
\end{algorithm}

\begin{figure}[ht]
  \centering
  \includegraphics[scale=0.4]{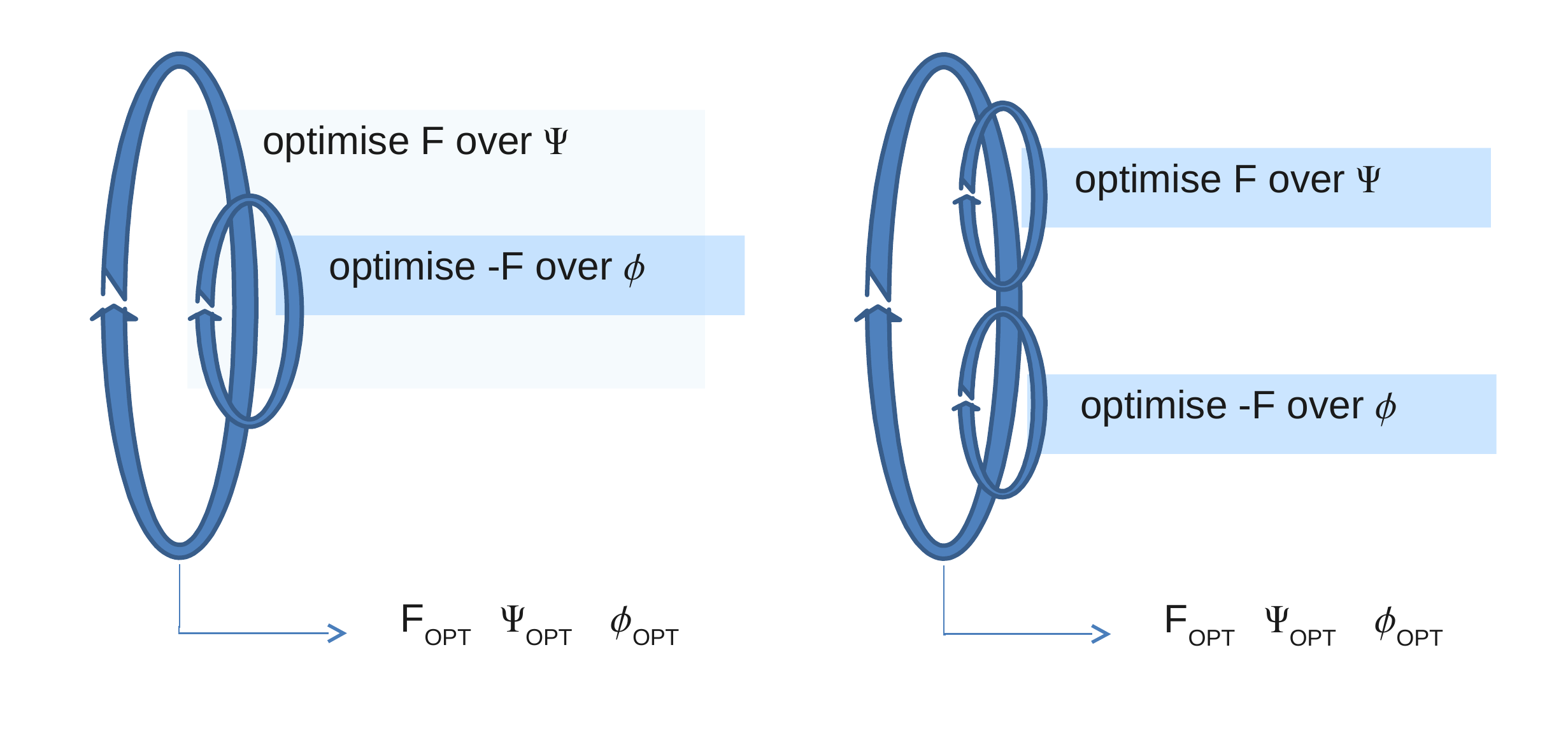}
  \caption{\label{fig:SchIt} {Schematic view of two iterative
      algorithms.} On the left-hand side, the nested optimization
    method studied in this paper optimizes the pseudo free-energy over
    one field for each optimization step taken for the other field.
    On the right hand side, the other iterative method puts both
    optimization loops on the same level in a third encompassing loop.
  }
\end{figure}

It is to be noted that this method differs from an iterative scheme
where both search loops are on the same level but included in a
general loop. The schematic view of these two programs given in
Fig.~(\ref{fig:SchIt}) shows these differences.  The classical
iterative method was implemented and tested but no convergence could
be reached.  We thus focus on the nested optimization method.

\subsubsection*{Squared gradient}\label{subsec:deriv}

An alternative to the nested optimizations discused above is the
use of a new functional based on the gradient of the free energy
\cite{PolyEl03}.  At the saddle point
\begin{equation}
  \left( \frac{\delta F}{\delta \phi} \right )_{\phi_{SP},\Psi_{SP}} = 
  \left( \frac{\delta F}{\delta \Psi} \right )_{\phi_{SP},\Psi_{SP}} = 0 
\end{equation}
We can build a new functional which is always positive and vanishes at
the stationary point:
\begin{equation}
  F_{deriv} (\phi, \Psi) = \int   \left[
    \left( \frac{\delta F}{\delta \Psi} \right) ^2
    + \left( \frac{\delta F}{\delta \phi} \right) ^2 \right] \dr
  \label{eq:Grad-01}
\end{equation}
The minimum of this functional yields the fields at equilibrium. The
advantage of this formulation 
 is that all fields can be treated equally
by the minimization process which can be managed as a single global
optimization.

\subsubsection*{Generalized scalar functionals}

The next algorithm that we test is that based on
eq.~(\ref{eq:monica}).  We denote
$\xi(\phi,\Psi) = \frac{\epsilon \nabla^2 \phi +\sigma -
  pe \Psi^2 }{ e \lambda}$,  and work with the free energy 
\begin{align*}
    F_I(\phi,&\Psi) \: =  \int  \Big [ \: k_B T \left\lbrace \frac{a^2}{6}
      (\nabla\Psi)^2 +\frac{v}{2} \Psi ^4 \right\rbrace
    + \frac{\epsilon}{2} (\nabla \phi)^2 \nonumber \\
     + &2 \lambda k_B T \: \Big \lbrace - \cosh \left( \beta e \phi
    \right)
    \; +  \; \beta e \phi \; \sinh(\beta e \phi )\nonumber \\
      -& \; \ash(\xi/2) \; \sinh(\beta e \phi) + \; (\xi/2) \;
    \ash \left( \xi/2 \right) \Big \rbrace \Big ] \dr
\end{align*}
The global minimum is found with a single optimization loop.

\subsubsection*{Legendre Transform}

The Legendre transform of the electrostatic degrees of freedom is
still possible in the presence of the extra polymer field.  We find
the locally defined free energy:
\begin{align*}
  F_L(\D,\Psi) = \int &\Big[ k_B T \left\lbrace \frac{a^2}{6}
                        (\nabla\Psi)^2 +\frac{v}{2} \Psi ^4 \right\rbrace
  \\+ &\frac{\D^2}{2\epsilon}  + k_BT \lambda g(\xi)
        \Big ] \dr
        \label{eq:TL2}
\end{align*}
with $\xi ={(\sigma - pe \Psi^2 -\div\D)}/{ e \lambda} $.  The minimum
over both $\Psi$ and $\D$ is found with a single optimization loop.

\subsection*{Numerical Results}
\label{sec:NumRes}

\begin{figure}[ht]
  \includegraphics[scale=.6] {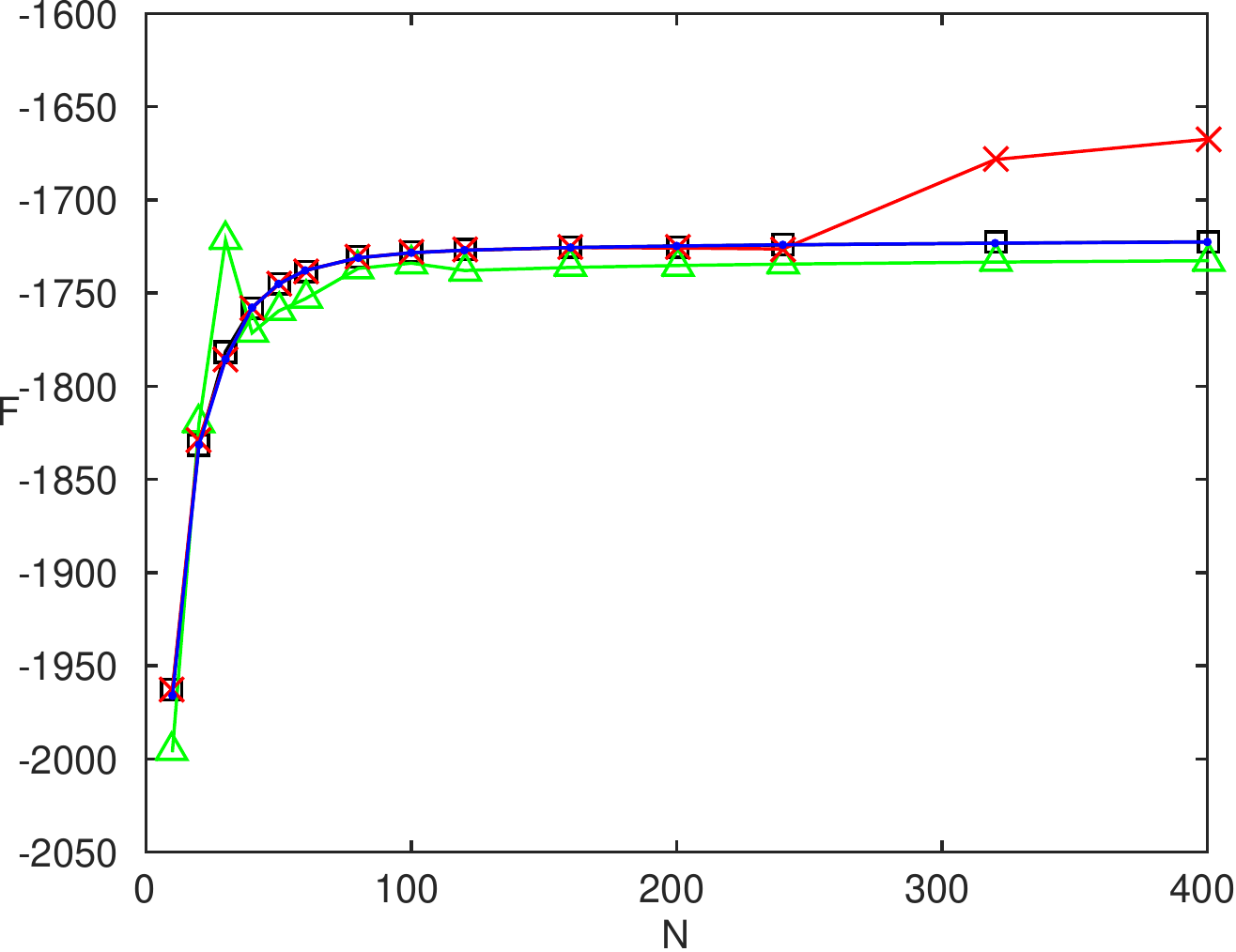}
  \caption{Convergence of free energy as a function of the number of
    points in the discretization. Blue dots ({\color{blue}$\bullet$}):
    Legendre transform. Green triangles:
    ({\color{green}$\vartriangle$}) squared gradient. Black squares:
    ($\square$) nested loops. Red crosses ({\color{red}$\times$}):
    generalized scalar functionals. Simple minimizer without external
    derivative information. Only the Legendre method reliably
    converges for all $N$.}\label{fig:virus}
\end{figure}

The four functionals were programmed in \textit{Matlab}. We use the
function \textit{fminunc} to search for the minimum for each
case. This function uses a quasi-Newton algorithm\cite{M04} and in
particular the Broyden-Fletche-Goldfarb-Shanno update. The idea is to
use a good, black-box minimizer to study how each formulation
converges in time and accuracy as the number of discretization points
increase. The ideal is a formulation that converges to high accuracy
without the need for detailed tuning of the algorithm for each
application. We choose to stop
iteration when the functional is converged to within an estimated
fraction of $10^{-12}$.

Three quantities are considered to study the performance of each
method:
\begin{itemize}
\item Convergence of $F$ when the number of points increase.
\item The derivative of the functional with respect to each field.  We
  use the $L_1$-norm to test the validity of the simulation
\item Simulation time as a function of $N$.
\end{itemize}

\begin{figure}[ht]
   \centering
   \includegraphics[scale=0.45]{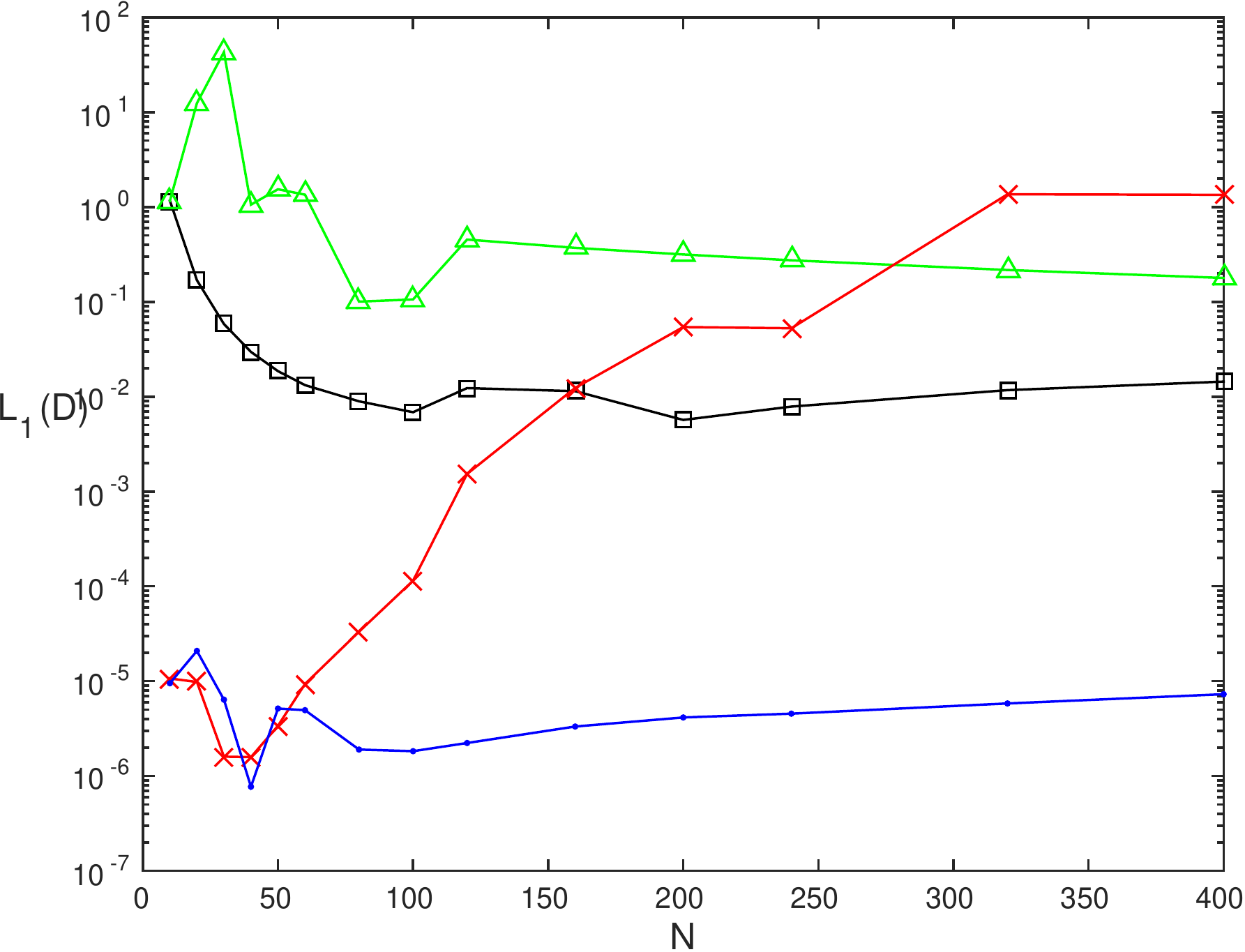}
   \caption{Color code as
       Fig.~\ref{fig:virus}. Derivatives of the free energy for the
       virus model for different discretizations and methods on exit
     from the optimization loop.}\label{fig:G_set005}
 \end{figure}

Variables are initialized to $\Psi=1$ for the polyelectrolyte field
and $\phi=-1$ for the potential.  The system size is $24$ nm, where
the charge of the capsid is \(\sigma = 0.4 e\)
at $R=12$ nm. The bulk concentration of monovalent ions is
\( \lambda = 10 \mbox{ mmol.L}^{-1}\)
and the water relative permittivity is $\epsilon_R=80$. This
corresponds to a Debye length of $3\text{ nm}$.  The parameters of the
polyelectrolyte are set to $a=0.5\text{ nm}$, $v = 0.05 \text{ nm}^3$,
and $p=1$.  The evolution of the free energies as a function of $N$ is
depicted in Fig.~(\ref{fig:virus}); Fig.~(\ref{fig:G_set005}) shows
the derivatives on stopping. 

 We work with values of $N$ up to 400 which gives values of
 ($1/\kappa h$), eq.~(\ref{eq:stiff}), up to 50. The discretizations
 are thus rather stiff.  The most robust method appears to be that
 based on the Legendre transform. Indeed, it is the { method that 
 converges and gives the smallest error} 
when the number of points is large, Fig.~(\ref{fig:virus}),
 Fig.~(\ref{fig:G_set005}). We note that the squared gradient
method gives poor results with even small $N$ 
 and the generalized scalar functional shows poor convergence for large $N$. 

\begin{figure}[ht]
  \centering
  \includegraphics[scale=0.45]{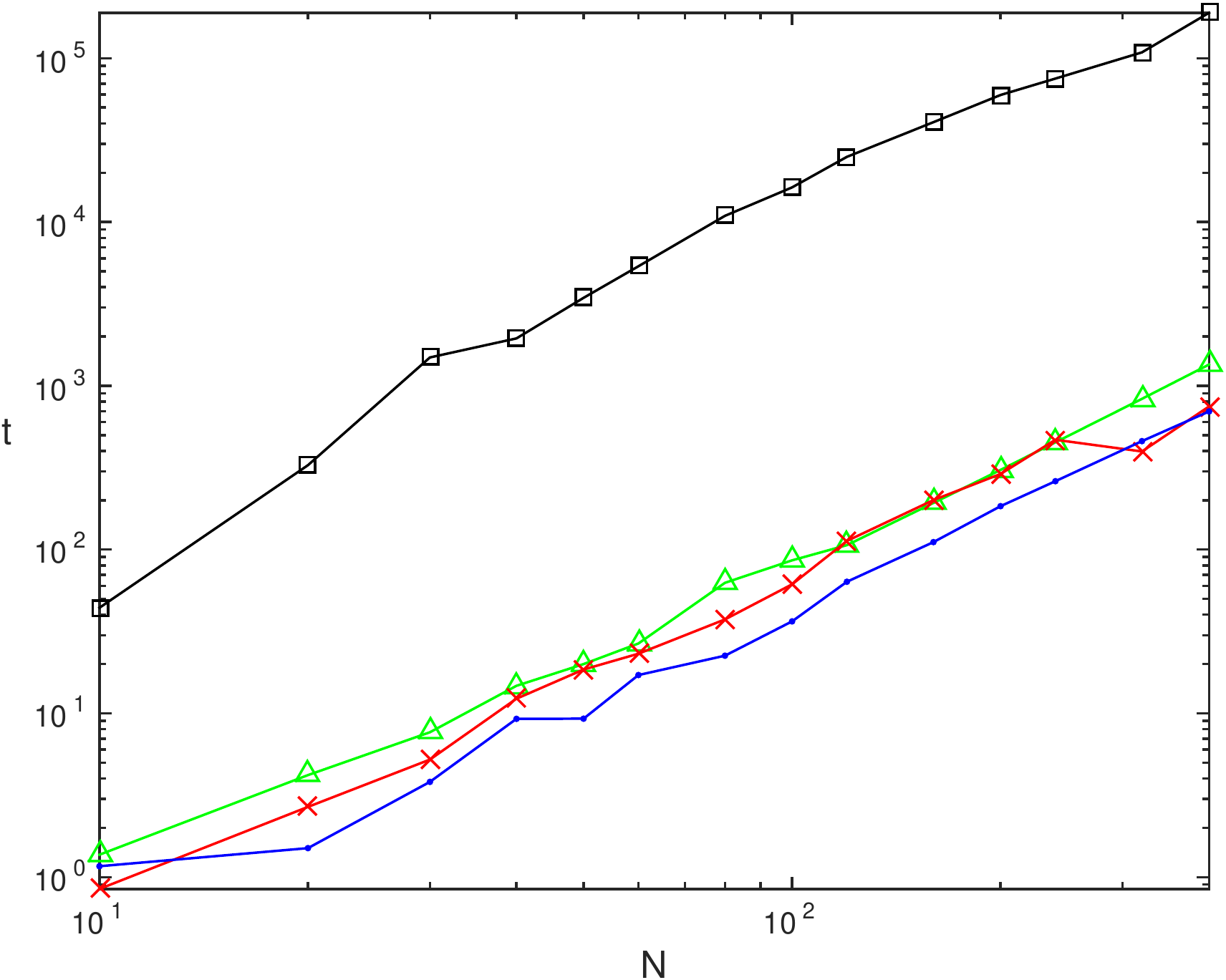}
  \caption{{Color code as
      Fig.~\ref{fig:virus}. Total simulation time as a function of the
      number of points in the discretization of the free energy. All
      algorithms display a quadratic growth in time with $N$. The
      algorithm based on the double loop is much slower than the three
      other methods.  Dash line slope 2.}}\label{fig:T_set005}
\end{figure}

 The  simulation time of each method is shown in
 Fig.~(\ref{fig:T_set005}) as a function of discretization; we use a
 recent intel-based desktop computer.  As
 expected the minimization time scales quadratically with the number
 of variables in the discretization. We tried other, random,
 initialisations and found that in all cases the method based on the
 Legendre transform gives the most stable results, together with a
 time of calculation which is as good as the other tested algorithms.

\subsection*{Stiffness and algorithms}

From our numerical experiments we see that naive optimization of an
expression with a saddle point leads to slow convergence. 
 It is more surprising that some
algorithms that can use convex optimization also give poor numerical
results. One possible explanation for the differences observed between
the three convex methods is that the algorithms have very different
stiffnesses when the number of discretization points increases. As
noted above an algorithm based on the Legendre transform has a
stiffness which increases as $N^2$, whereas both the squared gradient
and generalized scalar functionals become much stiffer with scaling in
$N^4$. This renders numerical codes much more susceptible to numerical
round-off errors and also requires more careful stopping criteria.
However as stated above our philosophy is that we are looking for free
energies that are stable and easy to use without large amounts of
algorithmic tuning from the user.

\begin{figure}[ht]
  \centering
  \includegraphics[scale=0.45]{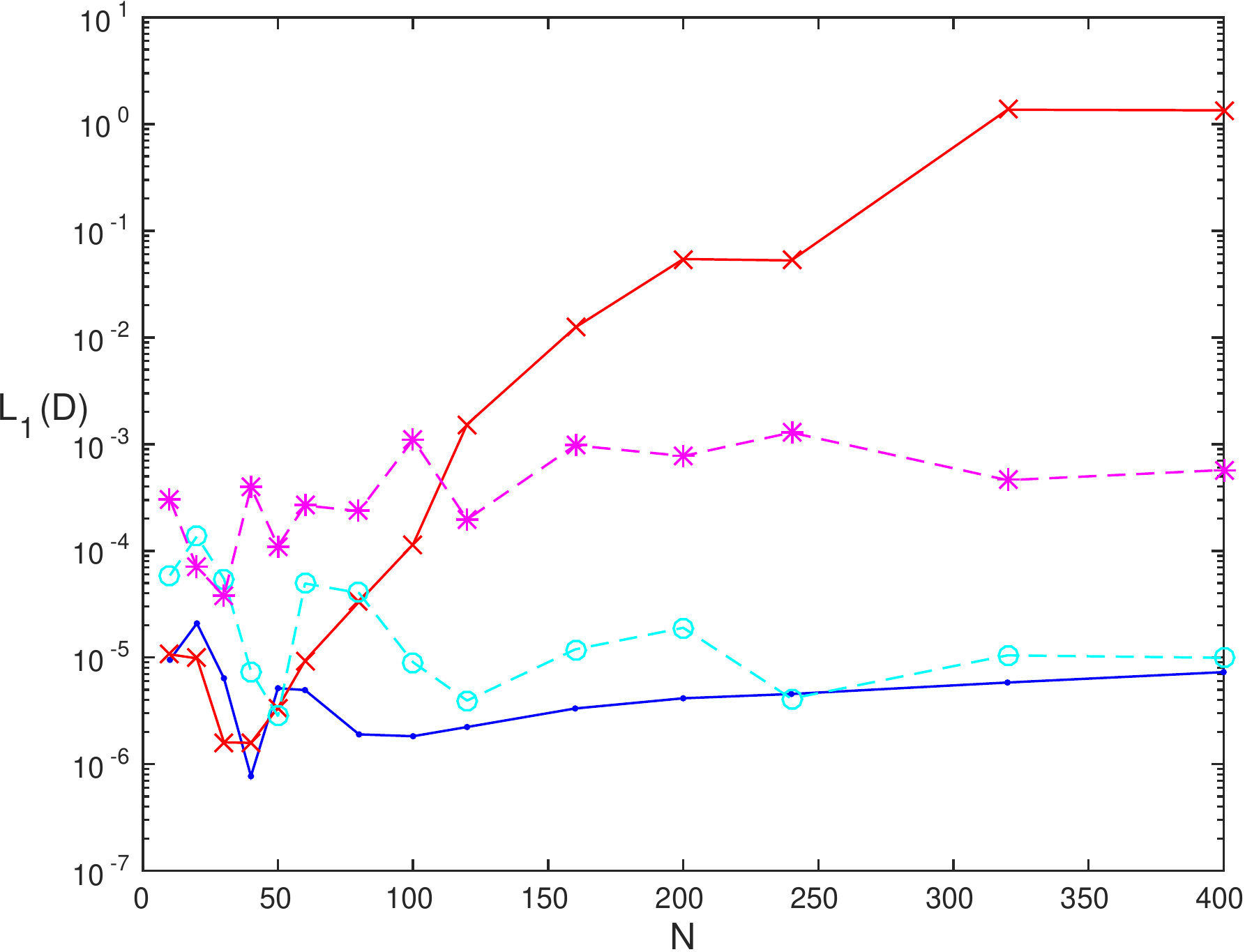}
  \caption{{Derivatives of the free energy for the
    virus model for different discretizations and algorithms on exit
    from the optimization loop.}
    Blue dots ({\color{blue}$\bullet$}): Legendre
    transform, quasi-Newton.  Cyan circles ({\color{cyan}$\circ$}):
    Legendre transform, trust-region. Red cross {\color{red}$\times$}:
    scalar functional, quasi-Newton.  Magenta stars
    ({\color{magenta}$\star$}): scalar functional,
    trust-region.}\label{fig:trustReg1}
\end{figure}

\begin{figure}[ht]
  \centering
  \includegraphics[scale=0.6]{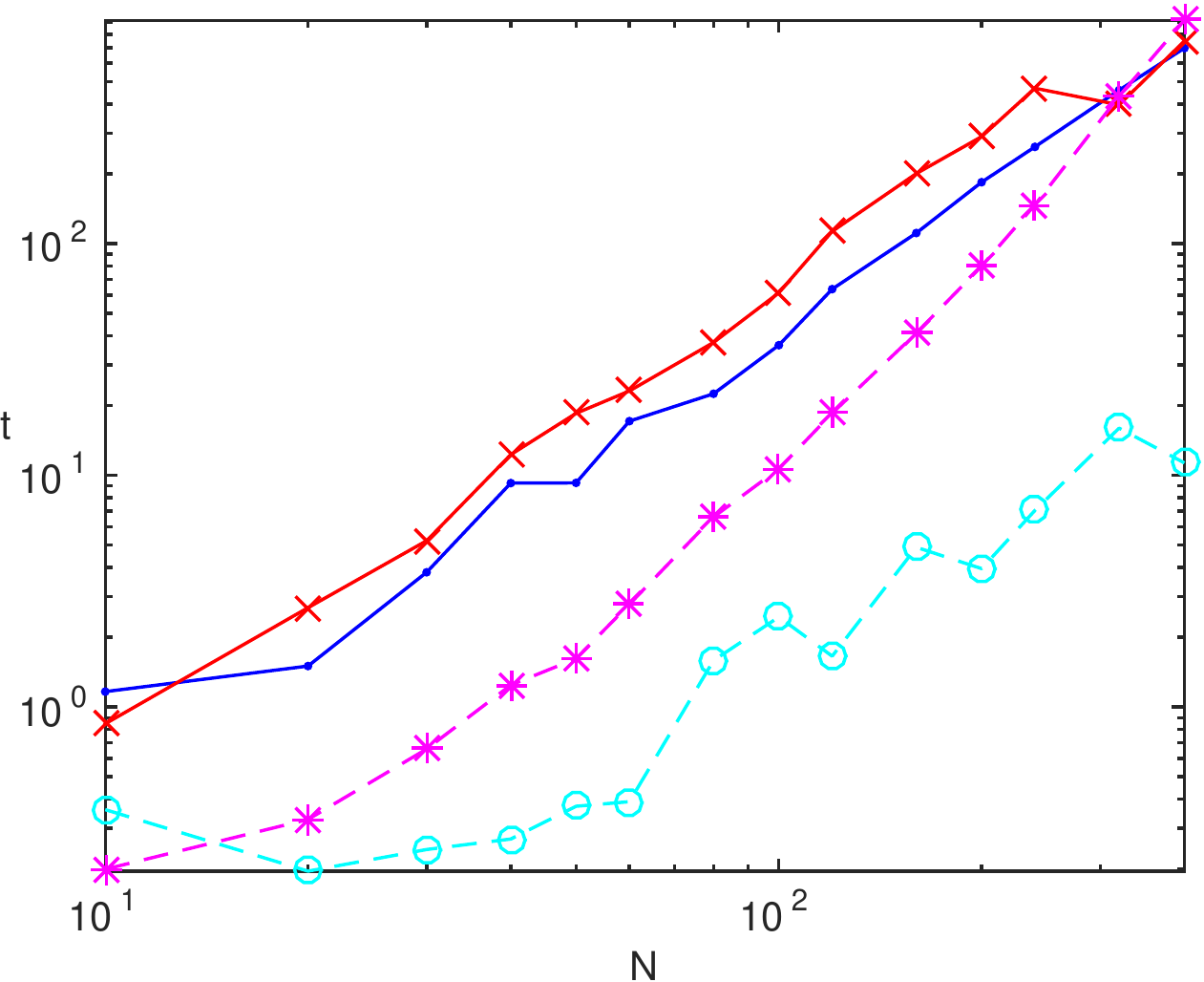}
  \caption{\label{fig:trustReg2} Color code as for
    Fig.~\ref{fig:trustReg1}. Computational time for different
    methods.}
\end{figure}

More sophisticated black-box algorithms are also available. In
particular if we are willing to calculate and program a routine which
calculates the first derivative of the free energy with respect to
each variable\footnote{as is required too in molecular dynamics; these
  derivatives are already calculated for the square gradient method}
we can find better results. In particular we use another algorithm
implemented in Matlab the {\it trust-region algorithm}\/ \cite{M05}.
It avoids over-large steps thanks to the limit imposed by the
definition of a trust-region, yet it maintains strong convergence
properties. The {L1-norm of the derivatives}
Fig.~(\ref{fig:trustReg1}) and time {used Fig.~(\ref{fig:trustReg2})}
are compared with the results obtained with the quasi-Newton
algorithm. The change of algorithm allows one either to significantly
reduce the computational time (Legendre transform method, blue and
cyan points), or to better converge (generalized scalar
functionals, red and magenta points. With the trust-region algorithm
the lack of convergence of the free energy disappears). Thus,
using the more sophisticated algorithm with derivative information
renders the optimization more reliable or more efficient.

\section*{Conclusion}

We have studied the quadratic fluctuations around equilibrium for
several expressions for the free energy of charged media.  These
fluctuations are closely related to dispersion energies and the
Debye-H\"uckel contribution to the free energy of electrolytes. The
functional based on the field $\D$ correctly reproduces this free
energy, even though the functional was originally proposed as a
minimum principle.

When used in molecular dynamics codes it is important to chose
functionals which are not too stiff. We emphasised that the ratio of
the largest to smallest eigenvalue in the quadratic form is closely
related to the number of time steps which are required to sample all
modes. Again the functional based on the vector field $\D$ seems to
display good scaling.

We used a model of a virus to implement and compare the optimization
of four free energy functionals. The Legendre-transformed functional
shows  faster convergence and a greater accuracy in our tests. We
suspect that this is linked to the lower stiffness of the Legendre
formulation of the free energy.

\subsection*{Acknowledgment}

Work financed in part by the ANR grant FSCF.
We wish to thank Derek Frydel for discussions 
and critical readings of the manuscript.

\section*{Appendix: Spherical symmetry and discretisation}

Our study of determinant identities brought out the importance of
discretizations that conserve the adjoint relation between the
divergence and gradient operators. Such identities are also important
for our study of the minimization of the polyelectrolyte free
energy. Non-compatible discretizations give rise to numerical errors
which are different between the different formulations. We now show
how to discretize within a spherical geometry in such a way as to
include important dualities between the discretized equations.  The potentials $\phi$
and $\Psi$ are defined on the points of the discretisation, while
$\nabla\phi$, $\nabla\Psi$ and $\D$ are defined on intermediate links.
We wish to conserve the fundamental identity
\begin{equation}
  \int \D \cdot \nabla \phi = - \int \phi \div \D + \oint \phi \D  \cdot dS
  \label{eq:SpSym01}
\end{equation}
The {discretized} left hand side of this equation gives:
\begin{equation}
  4 \pi \sum_n r_{n,n+1}^2 \left(\frac{\phi_{n+1} - \phi_n}{\delta}\right) \D_{n,n+1}
  \label{eq:SpSym02}
\end{equation}
where we have used the simplest differencing scheme for the gradient
operator and we define $r_{n,n+1}$ as the position of the centre of the link.
Identifying both sides of the identity, a definition of the
{discretized} divergence is obtained:
\begin{equation}
  \div_n \; \D = \frac{\D_{n,n+1} \: r_{n,n+1}^2 - \D_{n-1,n} \: r_{n-1,n}^2}{\delta \; r_n^2}
  \label{eq:SpSym03}
\end{equation}
The Laplacian operator is then defined as usual by $\nabla^2=-\div_n\, \grad$.

\section*{Appendix: one-loop free energies}

We give here the treatment of the one-loop correction in a homogeneous
electrolyte. The free energy (compared with the free energy of an
empty box) coming from quadratic fluctuations is
\begin{align}
F_f=& \frac{k_BT}{2} \sum_q \log 
 (1+ \frac{\kappa^2}{q^2}) \label{eq:oneint} \\
=& \frac{k_BT}{(2\pi)^2}  \int_0^{q_0} q^2 \dq \log(1 + \kappa^2/q^2)\\
=&  \frac{k_BT}{(2\pi)^2} \left (\kappa^2 q_0 -\frac{\pi \kappa^3}{3}
   \right ) +O(1/q_0)
\end{align}
Where we introduce an upper cut-off $q_0$. The divergence in $q_0^1$
can be balanced by a purely local self energy of the form $1/(2q^2)$
corresponding to the Born energy $1/8 \pi \epsilon a$ per atom, with
$a$ a real-space cut-off. The long-ranged Debye-H\"uckel energy to the
electrolyte is then
\begin{equation}
\frac{F_{DH}}{V} = - \frac{k_BT \kappa^3}{12 \pi}
\end{equation}
which comes from 
\begin{equation}
F_{DH} = \frac{k_BT}{2}\sum_q  \left (\log(1 + \kappa^2/q^2) -
  \lambda \beta e^2/(\epsilon q^2) \right) \label{eq:debyeIntegral}
\end{equation}

\bibliography{stiff}

\begin{thebibliography}{36}%
\makeatletter
\providecommand \@ifxundefined [1]{%
 \@ifx{#1\undefined}
}%
\providecommand \@ifnum [1]{%
 \ifnum #1\expandafter \@firstoftwo
 \else \expandafter \@secondoftwo
 \fi
}%
\providecommand \@ifx [1]{%
 \ifx #1\expandafter \@firstoftwo
 \else \expandafter \@secondoftwo
 \fi
}%
\providecommand \natexlab [1]{#1}%
\providecommand \enquote  [1]{``#1''}%
\providecommand \bibnamefont  [1]{#1}%
\providecommand \bibfnamefont [1]{#1}%
\providecommand \citenamefont [1]{#1}%
\providecommand \href@noop [0]{\@secondoftwo}%
\providecommand \href [0]{\begingroup \@sanitize@url \@href}%
\providecommand \@href[1]{\@@startlink{#1}\@@href}%
\providecommand \@@href[1]{\endgroup#1\@@endlink}%
\providecommand \@sanitize@url [0]{\catcode `\\12\catcode `\$12\catcode
  `\&12\catcode `\#12\catcode `\^12\catcode `\_12\catcode `\%12\relax}%
\providecommand \@@startlink[1]{}%
\providecommand \@@endlink[0]{}%
\providecommand \url  [0]{\begingroup\@sanitize@url \@url }%
\providecommand \@url [1]{\endgroup\@href {#1}{\urlprefix }}%
\providecommand \urlprefix  [0]{URL }%
\providecommand \Eprint [0]{\href }%
\providecommand \doibase [0]{http://dx.doi.org/}%
\providecommand \selectlanguage [0]{\@gobble}%
\providecommand \bibinfo  [0]{\@secondoftwo}%
\providecommand \bibfield  [0]{\@secondoftwo}%
\providecommand \translation [1]{[#1]}%
\providecommand \BibitemOpen [0]{}%
\providecommand \bibitemStop [0]{}%
\providecommand \bibitemNoStop [0]{.\EOS\space}%
\providecommand \EOS [0]{\spacefactor3000\relax}%
\providecommand \BibitemShut  [1]{\csname bibitem#1\endcsname}%
\let\auto@bib@innerbib\@empty
\bibitem [{\citenamefont {Allen}, \citenamefont {Hansen},\ and\ \citenamefont
  {Melchionna}(2001)}]{allen}%
  \BibitemOpen
  \bibfield  {author} {\bibinfo {author} {\bibfnamefont {R.}~\bibnamefont
  {Allen}}, \bibinfo {author} {\bibfnamefont {J.-P.}\ \bibnamefont {Hansen}}, \
  and\ \bibinfo {author} {\bibfnamefont {S.}~\bibnamefont {Melchionna}},\
  }\bibfield  {title} {\enquote {\bibinfo {title} {Electrostatic potential
  inside ionic solutions confined by dielectrics: a variational approach},}\
  }\href {\doibase 10.1039/B105176H} {\bibfield  {journal} {\bibinfo  {journal}
  {Phys. Chem. Chem. Phys.}\ }\textbf {\bibinfo {volume} {3}},\ \bibinfo
  {pages} {4177--4186} (\bibinfo {year} {2001})}\BibitemShut {NoStop}%
\bibitem [{\citenamefont {Fogolari}\ and\ \citenamefont {Briggs}(1997)}]{PB02}%
  \BibitemOpen
  \bibfield  {author} {\bibinfo {author} {\bibfnamefont {F.}~\bibnamefont
  {Fogolari}}\ and\ \bibinfo {author} {\bibfnamefont {J.~M.}\ \bibnamefont
  {Briggs}},\ }\bibfield  {title} {\enquote {\bibinfo {title} {On the
  variational approach to {Poisson-Boltzmann} free energies},}\ }\href
  {\doibase http://dx.doi.org/10.1016/S0009-2614(97)01193-7} {\bibfield
  {journal} {\bibinfo  {journal} {Chemical Physics Letters}\ }\textbf {\bibinfo
  {volume} {281}},\ \bibinfo {pages} {135 -- 139} (\bibinfo {year}
  {1997})}\BibitemShut {NoStop}%
\bibitem [{\citenamefont {Reiner}\ and\ \citenamefont {Radke}(1990)}]{reiner}%
  \BibitemOpen
  \bibfield  {author} {\bibinfo {author} {\bibfnamefont {E.~S.}\ \bibnamefont
  {Reiner}}\ and\ \bibinfo {author} {\bibfnamefont {C.~J.}\ \bibnamefont
  {Radke}},\ }\bibfield  {title} {\enquote {\bibinfo {title} {Variational
  approach to the electrostatic free energy in charged colloidal suspensions:
  general theory for open systems},}\ }\href@noop {} {\bibfield  {journal}
  {\bibinfo  {journal} {Journal of the Chemical Society, Faraday Transactions}\
  }\textbf {\bibinfo {volume} {86}},\ \bibinfo {pages} {3901--3912} (\bibinfo
  {year} {1990})}\BibitemShut {NoStop}%
\bibitem [{Note1()}]{Note1}%
  \BibitemOpen
  \bibinfo {note} {An exception is~\cite {vanderbilt}}\BibitemShut {NoStop}%
\bibitem [{\citenamefont {Landau}\ \emph {et~al.}(1984)\citenamefont {Landau},
  \citenamefont {Lifshits}, \citenamefont {Lifshits},\ and\ \citenamefont
  {Pitaevski{\u\i}}}]{landau}%
  \BibitemOpen
  \bibfield  {author} {\bibinfo {author} {\bibfnamefont {L.}~\bibnamefont
  {Landau}}, \bibinfo {author} {\bibfnamefont {E.}~\bibnamefont {Lifshits}},
  \bibinfo {author} {\bibfnamefont {E.}~\bibnamefont {Lifshits}}, \ and\
  \bibinfo {author} {\bibfnamefont {L.}~\bibnamefont {Pitaevski{\u\i}}},\
  }\href {http://books.google.fr/books?id=j7nvAAAAMAAJ} {\emph {\bibinfo
  {title} {Electrodynamics of continuous media:}}},\ Pergamon international
  library of science, technology, engineering, and social studies\ (\bibinfo
  {publisher} {Pergamon},\ \bibinfo {year} {1984})\BibitemShut {NoStop}%
\bibitem [{\citenamefont {Kirzhnits}(1987)}]{kirzh}%
  \BibitemOpen
  \bibfield  {author} {\bibinfo {author} {\bibfnamefont {D.~A.}\ \bibnamefont
  {Kirzhnits}},\ }\bibfield  {title} {\enquote {\bibinfo {title} {General
  properties of electromagnetic response functions},}\ }\href
  {http://stacks.iop.org/0038-5670/30/i=7/a=R02} {\bibfield  {journal}
  {\bibinfo  {journal} {Soviet Physics Uspekhi}\ }\textbf {\bibinfo {volume}
  {30}},\ \bibinfo {pages} {575} (\bibinfo {year} {1987})}\BibitemShut
  {NoStop}%
\bibitem [{\citenamefont {Dolgov}, \citenamefont {Kirzhnits},\ and\
  \citenamefont {Maksimov}(1981)}]{sign}%
  \BibitemOpen
  \bibfield  {author} {\bibinfo {author} {\bibfnamefont {O.}~\bibnamefont
  {Dolgov}}, \bibinfo {author} {\bibfnamefont {D.}~\bibnamefont {Kirzhnits}}, \
  and\ \bibinfo {author} {\bibfnamefont {E.}~\bibnamefont {Maksimov}},\
  }\bibfield  {title} {\enquote {\bibinfo {title} {On an admissible sign of the
  static dielectric function of matter},}\ }\href {\doibase
  10.1103/RevModPhys.53.81} {\bibfield  {journal} {\bibinfo  {journal} {Rev.
  Mod. Phys.}\ }\textbf {\bibinfo {volume} {53}},\ \bibinfo {pages} {81--93}
  (\bibinfo {year} {1981})}\BibitemShut {NoStop}%
\bibitem [{\citenamefont {Maggs}(2012)}]{Tony-LT}%
  \BibitemOpen
  \bibfield  {author} {\bibinfo {author} {\bibfnamefont {A.~C.}\ \bibnamefont
  {Maggs}},\ }\bibfield  {title} {\enquote {\bibinfo {title} {A minimizing
  principle for the poisson-boltzmann equation},}\ }\href
  {http://stacks.iop.org/0295-5075/98/i=1/a=16012} {\bibfield  {journal}
  {\bibinfo  {journal} {EPL (Europhysics Letters)}\ }\textbf {\bibinfo {volume}
  {98}},\ \bibinfo {pages} {16012} (\bibinfo {year} {2012})}\BibitemShut
  {NoStop}%
\bibitem [{\citenamefont {Pujos}\ and\ \citenamefont
  {Maggs}(2012)}]{pujos2012legendre}%
  \BibitemOpen
  \bibfield  {author} {\bibinfo {author} {\bibfnamefont {J.~S.}\ \bibnamefont
  {Pujos}}\ and\ \bibinfo {author} {\bibfnamefont {A.~C.}\ \bibnamefont
  {Maggs}},\ }\href {http://arxiv.org/abs/1211.6601} {\enquote {\bibinfo
  {title} {Legendre transforms for electrostatic energies},}\ } (\bibinfo
  {year} {2012}),\ \bibinfo {note} {cite arxiv:1211.6601 Comment: 7 pages;
  CECAM workshop: New Challenges in Electrostatics of Soft and Disordered
  Matter (May 7, 2012 to May 10, 2012):
  http://www.cecam.org/workshop-0-689.html}\BibitemShut {NoStop}%
\bibitem [{\citenamefont {Arnold}\ \emph {et~al.}(2013)\citenamefont {Arnold},
  \citenamefont {Breitsprecher}, \citenamefont {Fahrenberger}, \citenamefont
  {Kesselheim}, \citenamefont {Lenz},\ and\ \citenamefont {Holm}}]{holm}%
  \BibitemOpen
  \bibfield  {author} {\bibinfo {author} {\bibfnamefont {A.}~\bibnamefont
  {Arnold}}, \bibinfo {author} {\bibfnamefont {K.}~\bibnamefont
  {Breitsprecher}}, \bibinfo {author} {\bibfnamefont {F.}~\bibnamefont
  {Fahrenberger}}, \bibinfo {author} {\bibfnamefont {S.}~\bibnamefont
  {Kesselheim}}, \bibinfo {author} {\bibfnamefont {O.}~\bibnamefont {Lenz}}, \
  and\ \bibinfo {author} {\bibfnamefont {C.}~\bibnamefont {Holm}},\ }\bibfield
  {title} {\enquote {\bibinfo {title} {Efficient algorithms for electrostatic
  interactions including dielectric contrasts},}\ }\href {\doibase
  10.3390/e15114569} {\bibfield  {journal} {\bibinfo  {journal} {Entropy}\
  }\textbf {\bibinfo {volume} {15}},\ \bibinfo {pages} {4569--4588} (\bibinfo
  {year} {2013})}\BibitemShut {NoStop}%
\bibitem [{\citenamefont {Car}\ and\ \citenamefont {Parrinello}(1985)}]{car}%
  \BibitemOpen
  \bibfield  {author} {\bibinfo {author} {\bibfnamefont {R.}~\bibnamefont
  {Car}}\ and\ \bibinfo {author} {\bibfnamefont {M.}~\bibnamefont
  {Parrinello}},\ }\bibfield  {title} {\enquote {\bibinfo {title} {Unified
  approach for molecular dynamics and density-functional theory},}\ }\href
  {\doibase 10.1103/PhysRevLett.55.2471} {\bibfield  {journal} {\bibinfo
  {journal} {Phys. Rev. Lett.}\ }\textbf {\bibinfo {volume} {55}},\ \bibinfo
  {pages} {2471--2474} (\bibinfo {year} {1985})}\BibitemShut {NoStop}%
\bibitem [{\citenamefont {Rottler}\ and\ \citenamefont
  {Maggs}(2004)}]{rottler}%
  \BibitemOpen
  \bibfield  {author} {\bibinfo {author} {\bibfnamefont {J.}~\bibnamefont
  {Rottler}}\ and\ \bibinfo {author} {\bibfnamefont {A.~C.}\ \bibnamefont
  {Maggs}},\ }\bibfield  {title} {\enquote {\bibinfo {title} {Local molecular
  dynamics with {Coulombic} interactions},}\ }\href {\doibase
  10.1103/PhysRevLett.93.170201} {\bibfield  {journal} {\bibinfo  {journal}
  {Phys. Rev. Lett.}\ }\textbf {\bibinfo {volume} {93}},\ \bibinfo {pages}
  {170201} (\bibinfo {year} {2004})}\BibitemShut {NoStop}%
\bibitem [{\citenamefont {Jadhao}, \citenamefont {Solis},\ and\ \citenamefont
  {de~la Cruz}(2013)}]{PB01-Monica}%
  \BibitemOpen
  \bibfield  {author} {\bibinfo {author} {\bibfnamefont {V.}~\bibnamefont
  {Jadhao}}, \bibinfo {author} {\bibfnamefont {F.~J.}\ \bibnamefont {Solis}}, \
  and\ \bibinfo {author} {\bibfnamefont {M.~O.}\ \bibnamefont {de~la Cruz}},\
  }\bibfield  {title} {\enquote {\bibinfo {title} {Free-energy functionals of
  the electrostatic potential for {Poisson-Boltzmann} theory},}\ }\href
  {\doibase 10.1103/PhysRevE.88.022305} {\bibfield  {journal} {\bibinfo
  {journal} {Phys. Rev. E}\ }\textbf {\bibinfo {volume} {88}},\ \bibinfo
  {pages} {022305} (\bibinfo {year} {2013})}\BibitemShut {NoStop}%
\bibitem [{\citenamefont {Solis}, \citenamefont {Jadhao},\ and\ \citenamefont
  {Olvera de~la Cruz}(2013)}]{monica2}%
  \BibitemOpen
  \bibfield  {author} {\bibinfo {author} {\bibfnamefont {F.}~\bibnamefont
  {Solis}}, \bibinfo {author} {\bibfnamefont {V.}~\bibnamefont {Jadhao}}, \
  and\ \bibinfo {author} {\bibfnamefont {M.}~\bibnamefont {Olvera de~la
  Cruz}},\ }\bibfield  {title} {\enquote {\bibinfo {title} {Generating true
  minima in constrained variational formulations via modified {Lagrange}
  multipliers},}\ }\href {\doibase 10.1103/PhysRevE.88.053306} {\bibfield
  {journal} {\bibinfo  {journal} {Phys. Rev. E}\ }\textbf {\bibinfo {volume}
  {88}},\ \bibinfo {pages} {053306} (\bibinfo {year} {2013})}\BibitemShut
  {NoStop}%
\bibitem [{\citenamefont {Jadhao}, \citenamefont {Solis},\ and\ \citenamefont
  {de~la Cruz}(2012)}]{monica3}%
  \BibitemOpen
  \bibfield  {author} {\bibinfo {author} {\bibfnamefont {V.}~\bibnamefont
  {Jadhao}}, \bibinfo {author} {\bibfnamefont {F.}~\bibnamefont {Solis}}, \
  and\ \bibinfo {author} {\bibfnamefont {M.}~\bibnamefont {de~la Cruz}},\
  }\bibfield  {title} {\enquote {\bibinfo {title} {Simulation of charged
  systems in heterogeneous dielectric media via a true energy functional},}\
  }\href {\doibase 10.1103/PhysRevLett.109.223905} {\bibfield  {journal}
  {\bibinfo  {journal} {Phys. Rev. Lett.}\ }\textbf {\bibinfo {volume} {109}},\
  \bibinfo {pages} {223905} (\bibinfo {year} {2012})}\BibitemShut {NoStop}%
\bibitem [{\citenamefont {Maggs}(2004)}]{auxiliary}%
  \BibitemOpen
  \bibfield  {author} {\bibinfo {author} {\bibfnamefont {A.~C.}\ \bibnamefont
  {Maggs}},\ }\bibfield  {title} {\enquote {\bibinfo {title} {Auxiliary field
  {Monte Carlo} for charged particles},}\ }\href {\doibase
  http://dx.doi.org/10.1063/1.1642587} {\bibfield  {journal} {\bibinfo
  {journal} {The Journal of Chemical Physics}\ }\textbf {\bibinfo {volume}
  {120}},\ \bibinfo {pages} {3108--3118} (\bibinfo {year} {2004})}\BibitemShut
  {NoStop}%
\bibitem [{\citenamefont {Maggs}(2002)}]{dynamics}%
  \BibitemOpen
  \bibfield  {author} {\bibinfo {author} {\bibfnamefont {A.~C.}\ \bibnamefont
  {Maggs}},\ }\bibfield  {title} {\enquote {\bibinfo {title} {Dynamics of a
  local algorithm for simulating {Coulomb} interactions},}\ }\href {\doibase
  http://dx.doi.org/10.1063/1.1487821} {\bibfield  {journal} {\bibinfo
  {journal} {The Journal of Chemical Physics}\ }\textbf {\bibinfo {volume}
  {117}},\ \bibinfo {pages} {1975--1981} (\bibinfo {year} {2002})}\BibitemShut
  {NoStop}%
\bibitem [{\citenamefont {Pasquali}\ and\ \citenamefont {Maggs}(2008)}]{sam}%
  \BibitemOpen
  \bibfield  {author} {\bibinfo {author} {\bibfnamefont {S.}~\bibnamefont
  {Pasquali}}\ and\ \bibinfo {author} {\bibfnamefont {A.~C.}\ \bibnamefont
  {Maggs}},\ }\bibfield  {title} {\enquote {\bibinfo {title}
  {Fluctuation-induced interactions between dielectrics in general
  geometries},}\ }\href {\doibase http://dx.doi.org/10.1063/1.2949508}
  {\bibfield  {journal} {\bibinfo  {journal} {The Journal of Chemical Physics}\
  }\textbf {\bibinfo {volume} {129}},\ \bibinfo {eid} {014703} (\bibinfo {year}
  {2008})}\BibitemShut {NoStop}%
\bibitem [{\citenamefont {Podgornik}\ and\ \citenamefont {Zeks}(1988)}]{zeks}%
  \BibitemOpen
  \bibfield  {author} {\bibinfo {author} {\bibfnamefont {R.}~\bibnamefont
  {Podgornik}}\ and\ \bibinfo {author} {\bibfnamefont {B.}~\bibnamefont
  {Zeks}},\ }\bibfield  {title} {\enquote {\bibinfo {title} {Inhomogeneous
  coulomb fluid. a functional integral approach},}\ }\href {\doibase
  10.1039/F29888400611} {\bibfield  {journal} {\bibinfo  {journal} {J. Chem.
  Soc.{,} Faraday Trans. 2}\ }\textbf {\bibinfo {volume} {84}},\ \bibinfo
  {pages} {611--631} (\bibinfo {year} {1988})}\BibitemShut {NoStop}%
\bibitem [{\citenamefont {Netz}\ and\ \citenamefont {Orland}(1999)}]{henri}%
  \BibitemOpen
  \bibfield  {author} {\bibinfo {author} {\bibfnamefont {R.~R.}\ \bibnamefont
  {Netz}}\ and\ \bibinfo {author} {\bibfnamefont {H.}~\bibnamefont {Orland}},\
  }\bibfield  {title} {\enquote {\bibinfo {title} {Field theory for charged
  fluids and colloids},}\ }\href {http://stacks.iop.org/0295-5075/45/i=6/a=726}
  {\bibfield  {journal} {\bibinfo  {journal} {EPL (Europhysics Letters)}\
  }\textbf {\bibinfo {volume} {45}},\ \bibinfo {pages} {726} (\bibinfo {year}
  {1999})}\BibitemShut {NoStop}%
\bibitem [{\citenamefont {Wang}(2010)}]{wang}%
  \BibitemOpen
  \bibfield  {author} {\bibinfo {author} {\bibfnamefont {Z.-G.}\ \bibnamefont
  {Wang}},\ }\bibfield  {title} {\enquote {\bibinfo {title} {Fluctuation in
  electrolyte solutions: The self energy},}\ }\href {\doibase
  10.1103/PhysRevE.81.021501} {\bibfield  {journal} {\bibinfo  {journal} {Phys.
  Rev. E}\ }\textbf {\bibinfo {volume} {81}},\ \bibinfo {pages} {021501}
  (\bibinfo {year} {2010})}\BibitemShut {NoStop}%
\bibitem [{\citenamefont {Zia}, \citenamefont {Redish},\ and\ \citenamefont
  {McKay}(2009)}]{zia}%
  \BibitemOpen
  \bibfield  {author} {\bibinfo {author} {\bibfnamefont {R.~K.~P.}\
  \bibnamefont {Zia}}, \bibinfo {author} {\bibfnamefont {E.~F.}\ \bibnamefont
  {Redish}}, \ and\ \bibinfo {author} {\bibfnamefont {S.~R.}\ \bibnamefont
  {McKay}},\ }\bibfield  {title} {\enquote {\bibinfo {title} {{Making sense of
  the {L}egendre transform}},}\ }\href@noop {} {\bibfield  {journal} {\bibinfo
  {journal} {American Journal of Physics}\ }\textbf {\bibinfo {volume} {77}},\
  \bibinfo {pages} {614--622} (\bibinfo {year} {2009})}\BibitemShut {NoStop}%
\bibitem [{\citenamefont {Buyukdagli}, \citenamefont {Achim},\ and\
  \citenamefont {Ala-Nissila}(2012)}]{sahin}%
  \BibitemOpen
  \bibfield  {author} {\bibinfo {author} {\bibfnamefont {S.}~\bibnamefont
  {Buyukdagli}}, \bibinfo {author} {\bibfnamefont {C.~V.}\ \bibnamefont
  {Achim}}, \ and\ \bibinfo {author} {\bibfnamefont {T.}~\bibnamefont
  {Ala-Nissila}},\ }\bibfield  {title} {\enquote {\bibinfo {title}
  {Electrostatic correlations in inhomogeneous charged fluids beyond loop
  expansion},}\ }\href@noop {} {\bibfield  {journal} {\bibinfo  {journal} {J.
  Chem. Phys.}\ }\textbf {\bibinfo {volume} {137}},\ \bibinfo {pages} {104902}
  (\bibinfo {year} {2012})}\BibitemShut {NoStop}%
\bibitem [{\citenamefont {Buyukdagli}\ and\ \citenamefont
  {Ala-Nissila}(2014)}]{sahin2}%
  \BibitemOpen
  \bibfield  {author} {\bibinfo {author} {\bibfnamefont {S.}~\bibnamefont
  {Buyukdagli}}\ and\ \bibinfo {author} {\bibfnamefont {T.}~\bibnamefont
  {Ala-Nissila}},\ }\bibfield  {title} {\enquote {\bibinfo {title}
  {Electrostatic correlations on the ionic selectivity of cylindrical membrane
  nanopores},}\ }\href {\doibase http://dx.doi.org/10.1063/1.4864323}
  {\bibfield  {journal} {\bibinfo  {journal} {The Journal of Chemical Physics}\
  }\textbf {\bibinfo {volume} {140}},\ \bibinfo {eid} {064701} (\bibinfo {year}
  {2014})}\BibitemShut {NoStop}%
\bibitem [{\citenamefont {Naji}\ \emph {et~al.}(2013)\citenamefont {Naji},
  \citenamefont {Kanduč}, \citenamefont {Forsman},\ and\ \citenamefont
  {Podgornik}}]{rudireview}%
  \BibitemOpen
  \bibfield  {author} {\bibinfo {author} {\bibfnamefont {A.}~\bibnamefont
  {Naji}}, \bibinfo {author} {\bibfnamefont {M.}~\bibnamefont {Kanduč}},
  \bibinfo {author} {\bibfnamefont {J.}~\bibnamefont {Forsman}}, \ and\
  \bibinfo {author} {\bibfnamefont {R.}~\bibnamefont {Podgornik}},\ }\bibfield
  {title} {\enquote {\bibinfo {title} {Perspective: Coulomb fluids—weak
  coupling, strong coupling, in between and beyond},}\ }\href {\doibase
  http://dx.doi.org/10.1063/1.4824681} {\bibfield  {journal} {\bibinfo
  {journal} {The Journal of Chemical Physics}\ }\textbf {\bibinfo {volume}
  {139}},\ \bibinfo {eid} {150901} (\bibinfo {year} {2013})}\BibitemShut
  {NoStop}%
\bibitem [{\citenamefont {Xu}\ and\ \citenamefont {Maggs}(2014)}]{bPB04}%
  \BibitemOpen
  \bibfield  {author} {\bibinfo {author} {\bibfnamefont {Z.}~\bibnamefont
  {Xu}}\ and\ \bibinfo {author} {\bibfnamefont {A.}~\bibnamefont {Maggs}},\
  }\bibfield  {title} {\enquote {\bibinfo {title} {Solving fluctuation-enhanced
  {Poisson-Boltzmann} equations},}\ }\href {\doibase
  http://dx.doi.org/10.1016/j.jcp.2014.07.004} {\bibfield  {journal} {\bibinfo
  {journal} {Journal of Computational Physics}\ }\textbf {\bibinfo {volume}
  {275}},\ \bibinfo {pages} {310 -- 322} (\bibinfo {year} {2014})}\BibitemShut
  {NoStop}%
\bibitem [{\citenamefont {Koehl}\ and\ \citenamefont {Delarue}(2010)}]{koehl}%
  \BibitemOpen
  \bibfield  {author} {\bibinfo {author} {\bibfnamefont {P.}~\bibnamefont
  {Koehl}}\ and\ \bibinfo {author} {\bibfnamefont {M.}~\bibnamefont
  {Delarue}},\ }\bibfield  {title} {\enquote {\bibinfo {title} {Aquasol: An
  efficient solver for the dipolar {Poisson--Boltzmann--Langevin} equation},}\
  }\href {\doibase 10.1063/1.3298862} {\bibfield  {journal} {\bibinfo
  {journal} {The Journal of Chemical Physics}\ }\textbf {\bibinfo {volume}
  {132}},\ \bibinfo {eid} {064101} (\bibinfo {year} {2010})}\BibitemShut
  {NoStop}%
\bibitem [{\citenamefont {Lu}\ \emph {et~al.}(2008)\citenamefont {Lu},
  \citenamefont {Zhou}, \citenamefont {Holst},\ and\ \citenamefont
  {McCammon}}]{PB03}%
  \BibitemOpen
  \bibfield  {author} {\bibinfo {author} {\bibfnamefont {B.}~\bibnamefont
  {Lu}}, \bibinfo {author} {\bibfnamefont {Y.}~\bibnamefont {Zhou}}, \bibinfo
  {author} {\bibfnamefont {M.}~\bibnamefont {Holst}}, \ and\ \bibinfo {author}
  {\bibfnamefont {J.}~\bibnamefont {McCammon}},\ }\bibfield  {title} {\enquote
  {\bibinfo {title} {Recent progress in numerical methods for the
  {Poisson--Boltzmann} equation in biophysical applications},}\ }\href@noop {}
  {\bibfield  {journal} {\bibinfo  {journal} {Communications in Computational
  Physics}\ }\textbf {\bibinfo {volume} {3}},\ \bibinfo {pages} {973--1009}
  (\bibinfo {year} {2008})}\BibitemShut {NoStop}%
\bibitem [{\citenamefont {Edwards}(1965)}]{PolyEl08}%
  \BibitemOpen
  \bibfield  {author} {\bibinfo {author} {\bibfnamefont {S.~F.}\ \bibnamefont
  {Edwards}},\ }\bibfield  {title} {\enquote {\bibinfo {title} {The statistical
  mechanics of polymers with excluded volume},}\ }\href
  {http://stacks.iop.org/0370-1328/85/i=4/a=301} {\bibfield  {journal}
  {\bibinfo  {journal} {Proceedings of the Physical Society}\ }\textbf
  {\bibinfo {volume} {85}},\ \bibinfo {pages} {613} (\bibinfo {year}
  {1965})}\BibitemShut {NoStop}%
\bibitem [{\citenamefont {de~Gennes}(1979)}]{PolyEl07-PGG}%
  \BibitemOpen
  \bibfield  {author} {\bibinfo {author} {\bibfnamefont {P.-G.}\ \bibnamefont
  {de~Gennes}},\ }\enquote {\bibinfo {title} {Scaling concepts in polymer
  physics},}\ \ (\bibinfo  {publisher} {Cornell University Press},\ \bibinfo
  {year} {1979})\ Chap.\ \bibinfo {chapter} {IX.2. Self-Consistent
  Fields}\BibitemShut {NoStop}%
\bibitem [{\citenamefont {\ifmmode~\check{S}\else \v{S}\fi{}iber}\ and\
  \citenamefont {Podgornik}(2008)}]{Vir01}%
  \BibitemOpen
  \bibfield  {author} {\bibinfo {author} {\bibfnamefont {A.}~\bibnamefont
  {\ifmmode~\check{S}\else \v{S}\fi{}iber}}\ and\ \bibinfo {author}
  {\bibfnamefont {R.}~\bibnamefont {Podgornik}},\ }\bibfield  {title} {\enquote
  {\bibinfo {title} {Nonspecific interactions in spontaneous assembly of empty
  versus functional single-stranded rna viruses},}\ }\href {\doibase
  10.1103/PhysRevE.78.051915} {\bibfield  {journal} {\bibinfo  {journal} {Phys.
  Rev. E}\ }\textbf {\bibinfo {volume} {78}},\ \bibinfo {pages} {051915}
  (\bibinfo {year} {2008})}\BibitemShut {NoStop}%
\bibitem [{\citenamefont {Siber}, \citenamefont {Bozic},\ and\ \citenamefont
  {Podgornik}(2012)}]{Vir02}%
  \BibitemOpen
  \bibfield  {author} {\bibinfo {author} {\bibfnamefont {A.}~\bibnamefont
  {Siber}}, \bibinfo {author} {\bibfnamefont {A.~L.}\ \bibnamefont {Bozic}}, \
  and\ \bibinfo {author} {\bibfnamefont {R.}~\bibnamefont {Podgornik}},\
  }\bibfield  {title} {\enquote {\bibinfo {title} {Energies and pressures in
  viruses: contribution of nonspecific electrostatic interactions},}\ }\href
  {\doibase 10.1039/C1CP22756D} {\bibfield  {journal} {\bibinfo  {journal}
  {Phys. Chem. Chem. Phys.}\ }\textbf {\bibinfo {volume} {14}},\ \bibinfo
  {pages} {3746--3765} (\bibinfo {year} {2012})}\BibitemShut {NoStop}%
\bibitem [{\citenamefont {Borukhov}, \citenamefont {Andelman},\ and\
  \citenamefont {Orland}(1999)}]{PolyEl03}%
  \BibitemOpen
  \bibfield  {author} {\bibinfo {author} {\bibfnamefont {I.}~\bibnamefont
  {Borukhov}}, \bibinfo {author} {\bibfnamefont {D.}~\bibnamefont {Andelman}},
  \ and\ \bibinfo {author} {\bibfnamefont {H.}~\bibnamefont {Orland}},\
  }\bibfield  {title} {\enquote {\bibinfo {title} {Effect of polyelectrolyte
  adsorption on intercolloidal forces},}\ }\href {\doibase 10.1021/jp990055r}
  {\bibfield  {journal} {\bibinfo  {journal} {The Journal of Physical Chemistry
  B}\ }\textbf {\bibinfo {volume} {103}},\ \bibinfo {pages} {5042--5057}
  (\bibinfo {year} {1999})},\ \Eprint
  {http://arxiv.org/abs/http://pubs.acs.org/doi/pdf/10.1021/jp990055r}
  {http://pubs.acs.org/doi/pdf/10.1021/jp990055r} \BibitemShut {NoStop}%
\bibitem [{\citenamefont {Dennis}\ and\ \citenamefont {More}(1977)}]{M04}%
  \BibitemOpen
  \bibfield  {author} {\bibinfo {author} {\bibfnamefont {J.}~\bibnamefont
  {Dennis}, \bibfnamefont {J.~E.}}\ and\ \bibinfo {author} {\bibfnamefont
  {J.~J.}\ \bibnamefont {More}},\ }\bibfield  {title} {\enquote {\bibinfo
  {title} {{Quasi-Newton} methods, motivation and theory},}\ }\href@noop {}
  {\bibfield  {journal} {\bibinfo  {journal} {SIAM Review}\ }\textbf {\bibinfo
  {volume} {19}},\ \bibinfo {pages} {pp. 46--89} (\bibinfo {year}
  {1977})}\BibitemShut {NoStop}%
\bibitem [{Note2()}]{Note2}%
  \BibitemOpen
  \bibinfo {note} {As is required too in molecular dynamics; these derivatives
  are already calculated for the square gradient method}\BibitemShut {NoStop}%
\bibitem [{\citenamefont {Sorensen}(1982)}]{M05}%
  \BibitemOpen
  \bibfield  {author} {\bibinfo {author} {\bibfnamefont {D.~C.}\ \bibnamefont
  {Sorensen}},\ }\bibfield  {title} {\enquote {\bibinfo {title} {Newton's
  method with a model trust region modification},}\ }\href
  {http://www.jstor.org/stable/2156955} {\bibfield  {journal} {\bibinfo
  {journal} {SIAM Journal on Numerical Analysis}\ }\textbf {\bibinfo {volume}
  {19}},\ \bibinfo {pages} {pp. 409--426} (\bibinfo {year} {1982})}\BibitemShut
  {NoStop}%
\end{thebibliography}%

\end{document}